\documentclass[a4paper,11pt]{article}
\pdfoutput=1 
\usepackage{jheppub} 
\usepackage[T1]{fontenc}
\DeclareUnicodeCharacter{2009}{\,} 
\usepackage{slashed}
\usepackage{physics}

\usepackage{xcolor}

\newcommand{\Lmt}{$U(1)_{L_{\mu}-L_{\tau}}\ $}
\def\gsim{\:\raisebox{-0.5ex}{$\stackrel{\textstyle>}{\sim}$}\:}
\def\lsim{\:\raisebox{-0.5ex}{$\stackrel{\textstyle<}{\sim}$}\:}

\title{\Lmt for Light Dark Matter, $g_\mu - 2$, the $511$ keV excess and the Hubble Tension}
\author[a]{Manuel Drees}
\author[a]{Wenbin Zhao}
\affiliation[a]{Bethe Center for Theoretical Physics and Physikalisches Institut, Universität Bonn\\ Nussallee 12, 53115 Bonn, Germany}
\emailAdd{drees@th.physik.uni-bonn.de}
\emailAdd{wenbin.zhao@uni-bonn.de}

\abstract{ In this paper we introduce a light Dirac particle $\psi$ as thermal dark matter candidate in a \Lmt model. Together with the new gauge boson $X$, we find a possible parameter space with $m_X \simeq 20$ MeV, \Lmt coupling $g_X \simeq 5 \cdot 10^{-4}$ and $m_\psi \gsim m_X/2$ where the $(g-2)_\mu$ anomaly, dark matter, the Hubble tension, and (part of) the excess of $511$ keV photons from the region near the galactic center can be explained simultaneously. This model is safe from current experimental and astrophysical constraints, but can be probed by the next generation of neutrino experiments as well as low--energy $e^+e^-$ colliders.}
\makeatletter
\gdef\@fpheader{}
\makeatother
\begin{document}
\maketitle
\flushbottom

\section{Introduction}

The \Lmt model \cite{Intro:Z'} is a well--motivated extension of the
Standard Model (SM). It contains a new gauge boson which couples to
second and third generation leptons. The model is free of anomalies
with SM particle content; this remains true in the presence of
right--handed SM singlet neutrinos, if two of them carry equal but
opposite \Lmt charges of $\pm 1$. Moreover, it allows to explain
\cite{Fayet:2007ua, Pospelov:2008zw} the $4.2 \sigma$ discrepancy
between the recent measurement of the muon's anomalous magnetic moment
\cite{g-2:Fermilab} and SM predictions \cite{g-2:SM_value}.

The new gauge boson can also serve as a mediator between SM particles
and cosmological dark matter (DM) \cite{Fayet:2007ua}. Thus this model
has been used to study dark matter physics in the GeV to TeV range by
introducing spinor or scalar DM, as possible weakly interacting
massive particle (WIMP) candidates \cite{Intro:DM3, Intro:DM2,
  Biswas:2016yan, Kamada:2018zxi}. It can produce the correct thermal
relic density in minimal cosmology without violating existing DM
search bounds. However, in this mass range a relatively large \Lmt
coupling $g_X$ is required, which leads to a host of constraints
\cite{Drees:2018hhs, Chun:2018ibr}.

The experimental constraints from direct DM searches are considerably
weaker in the MeV range. Moreover, the explanation of $g_\mu - 2$ now
requires a gauge coupling below $10^{-3}$. Previous studies showed
that near-resonant DM annihilation, co-annihilation, or a DM charge $\gg 1$ is needed
in order to reproduce the DM relic abundance in this region of
parameter space \cite{Intro:lightdm1, Intro:lightdm2, Intro:lightdm3,DarkMatter:Asai_charge,lightdm:co-annihilation_Okada}.

For DM and gauge boson masses below the muon mass the dark sector
couples primarily to neutrinos. The decoupling of DM particles, and
the decay of the gauge bosons, can therefore increase the energy
density in neutrinos, corresponding to a change
$\delta N_{\text{eff}} \lsim 0.4$ \cite{Neff:Bosoncase}, which relaxes
the tension between cosmological and ``local'' measurements of the
Hubble constant \cite{Vagnozzi:2019ezj, DiValentino:2021izs}.

As already mentioned, the DM mass $m_\psi$ needs to be close to
$m_X/2$ in order to reproduce the thermal relic density in minimal
cosmology. If $m_\psi < m_X/2$ the thermally averaged DM annihilation
cross section $\langle \sigma v \rangle$ will fall after DM decoupling,
making DM annihilation in today's Universe essentially
unobservable. On the other hand, if $m_\psi > m_X/2$,
$\langle \sigma v \rangle$ will increase after DM decoupled. In this
case annihilation into $e^+e^-$ pairs, which is possible due to
kinetic mixing between the new gauge boson and the photon, can even
(help to) explain the excess of $511$ keV photons from the region
around the center of our galaxy that has been observed since the
1970's \cite{511:1970_1, 511:1970_2, 511:1970_3, 511:INTEGRAL, 511:COSI}. 

The remainder of this paper is organized as follows. In Sec.~2 we
introduce the model Lagrangian and discuss some relevant
phenomenology. In Sec.~3 we discuss important constraints on the new
gauge boson and DM particles from several different experiments and
observations. In Sec.~4 we apply our model to the 511 keV line, while
Sec.~5 briefly discusses future tests of the model. Finally, Sec.~6
contains a brief summary and some conclusions.

\section{The Model}

Our model is based on an
$SU(3)_c \times SU(2)_L \times U(1)_Y \times U(1)_{L_{\mu}-L_{\tau}}$
gauge theory. In addition to the well--known SM particles, we
introduce an SM singlet Dirac DM $\psi$ with \Lmt charge $q_\psi$ and
a spin 1 gauge boson $X$. The $\mu$ and $\tau$ family leptons are
assigned charge $1$ and $-1$, respectively. Since the Dirac DM particle
$\psi$ contains two two--component spinors with opposite \Lmt charges,
the model remains anomaly--free for any value of $q_\psi$.

The \Lmt model can also have a rich neutrino phenomenology
\cite{Neutrino:Asai_1, Neutrino:Asai_2, Neutrino:Asai_3}. We will
assume that the right--handed, SM singlet neutrinos needed to generate
realistic neutrino masses have masses well above $m_\psi$, in which
case they play no significant role in the phenomena studied in this
paper. Moreover, the Higgs boson needed to break \Lmt cannot couple to
$\psi$ directly and thus also plays no role in the phenomenology we are
interested in. The relevant part of the Lagrangian of our model thus
reads:
\begin{equation} \label{eq:L}
\begin{split}
\mathcal{L}=\mathcal{L}_{SM}&-g_XX_\lambda(\bar{\mu}\gamma^\lambda\mu-\bar{\tau}
\gamma^\lambda\tau+\bar{\nu}_{\mu L}\gamma^{\lambda}\nu_{\mu L}
    -\bar{\nu}_{\tau L}\gamma^{\lambda}\nu_{\tau L})\\
     &-\frac{1}{4}X_{\mu\nu}X^{\mu\nu}+\frac{1}{2}m_X^2X_{\mu}X^{\mu}\\
     &+\bar{\psi}(i\slashed{\partial}-m_{\psi})\psi -
     q_{\psi}g_XX_{\lambda}\bar{\psi}\gamma^{\lambda}\psi\,.
\end{split}
\end{equation}
Here $g_X$ is the \Lmt gauge coupling,
$X_{\mu \nu}= \partial_\mu X_{\nu}-\partial_\nu X_{\mu}$ is the \Lmt field
strength tensor and $m_X$ is the mass of the new gauge boson.

Note that we assume the \Lmt group to be orthogonal to the SM gauge group;
in particular, there is no tree--level kinetic mixing between the two $U(1)$
factors. However, once the vacuum expectation value of the SM Higgs boson
is taken into account, so that the charged leptons obtain masses, this
mixing is induced radiatively through loops involving $\mu$ and $\tau$ leptons.
This mixing adds an additional interaction term:
\begin{equation} \label{eq:mix}
\begin{split}
\mathcal{L}_{X,J_{\rm em}} &= - \epsilon_A e X_{\mu}J^{\mu}_{\rm em}\,,\\
\epsilon_A &= -\frac{eg_X}{12\pi^2} \ln{\left( \frac{m^2_\tau}{m^2_\mu}\right)}
\approx -\frac{g_X}{70}\,,
    \end{split}
\end{equation}
where $J_{\rm em}^\mu$ is the electromagnetic current. The mixing
between $X$ and the SM $Z$ boson is further suppressed by
$\frac{m_X^2}{m_Z^2}$; it is thus completely negligible for
$m_X \lsim 100$ MeV, which is the range of masses we are interested
in.

Important processes which determine dark matter phenomenon are
$\psi\bar{\psi}\to f\bar{f}$ annihilation, where $f$ are SM particles
with coupling to the $X$ boson. The corresponding cross section reads:
\begin{equation} \label{eq:sigma}
\sigma(s) = \frac{g_X^4 q_\psi^2}{12\pi}\sqrt{\frac{s-4m_f^2}{s-4m_\psi^2}}
\frac{s^2+2(m_\psi^2+m_f^2)s+4m_f^2m_\psi^2}{s[(s-m_X^2)^2+\Gamma_X^2m_X^2]}
\,.
\end{equation}
Here $\Gamma_X$ is the total width of the $X$ boson. It is the sum of
the partial widths:
\begin{equation} \label{eq:Gamma}
\begin{split}
  \Gamma(X \to l^+l^-) &= \frac{g_X^2m_X}{12\pi}(1+\frac{2m^2_l}{m_X^2})
  \sqrt{1-\frac{4m_l^2}{m_X^2}} \,;\\
  \Gamma(X \to \bar{\nu}_l\nu_l) &= \frac{g_X^2m_X}{24\pi}\,,
    \end{split}
\end{equation}
where $l = \mu$ or $\tau$.

Recently the Fermilab Muon $g-2$ Experiment has published their first
results, which confirmed a positive deviation from the SM prediction
\cite{g-2:Fermilab}. Together with older results from the BNL E821
experiment \cite{g-2:BNL}, the measurements exceed the SM prediction
\cite{g-2:SM_value} by $4.2 \ \sigma$:
\begin{equation} \label{eq:gm2exp}
  \Delta \alpha_\mu = \alpha_\mu(\text{Exp})-\alpha_\mu(\text{SM})
  = (251\pm 59)\times10^{-11}\,,
\end{equation}
which is a quite strong hint for BSM physics. Our model contributes to
$(g-2)_{\mu}$ at one--loop level \cite{Fayet:2007ua,Pospelov:2008zw}:
\begin{equation} \label{eq:gm2}
  \Delta \alpha_\mu = \frac{g_X^2}{8\pi^2} \int_0^1 dx
  \frac{2m^2_\mu x^2(1-x)}{x^2 m_\mu^2 + (1-x)m_X^2}\,.
\end{equation}
This reproduces the measurement (\ref{eq:gm2exp}) for
$g_X \simeq 4.4 \cdot 10^{-4}$ if $m_X^2 \ll m_\mu^2$, and for
$g_X \simeq 5.4 \cdot 10^{-4} m_X / m_\mu$ if $m_X^2 \gg m_\mu^2$.

\section{Constraints}

In this section we discuss the most relevant constraints on the
parameter space. We focus on cosmological and astrophysical constraints;
other constraints are briefly summarized in the last subsection.

\subsection{Dark Matter Relic Density}

We want $\psi$ to have the correct thermal relic density in standard
cosmology. We compute the relic density by approximately solving the
Boltzmann equation using the formalism established in
\cite{DM:three_expection_in_relic_abundances}:\footnote{We note in
  passing that {\tt MadDM} \cite{Ambrogi:2018jqj} showed serious
  numerical instabilities in parts of the parameter space with
  $m_\psi < m_X/2$.}
\begin{equation} \label{eq:boltz}
  \frac{dn_\psi}{dt} +3Hn_\psi = -\frac{\langle \sigma v \rangle }{2}
  (n^2_\psi-n^2_{\psi,eq})\,.
\end{equation}
We assume equal $\psi$ and $\bar\psi$ density; the factor $1/2$
appears since only $\psi \bar \psi$ encounters can lead to
annihilation, while $\psi \psi$ and $\bar \psi \bar \psi$ encounters
cannot.  For $m_\psi < m_\mu$ the most relevant annihilation process
is $\psi \bar{\psi} \to \nu_{\mu,\tau}\bar{\nu}_{\mu,\tau}$ mediated
by $X$ boson exchange in the $s-$channel. The corresponding thermally
averaged cross section is \cite{DM:ThermalaveragedCrossSection}:
\begin{equation} \label{eq:sigav}
\langle \sigma v \rangle =
\frac{1}{8m_\psi^4 T K_2^2(m_\psi/T)} \int_{4m_\psi^2}^\infty ds
\sigma(s) \sqrt{s} \left(s-4m^2_\psi\right)
K_1\left(\frac{\sqrt{s}}{T} \right)\,,
\end{equation}
where the $K_n$ are modified Bessel functions of order $n$ and the
annihilation cross section $\sigma(s)$ is given by eq.(\ref{eq:sigma})
with $m_f = 0$. The scaled inverse decoupling temperature
$x_f= \frac{m_\psi}{T_f}$ is determined by solving the following
equation iteratively:
\begin{equation} \label{eq:xf}
  x_f = \ln{\frac{0.076 M_{\rm Pl} m_\psi  \langle \sigma v \rangle}
    {g_*^{1/2}x_f^{1/2}}}\,.
\end{equation}
Here $M_{\rm Pl} = 1.22 \cdot 10^{19}$ GeV is the Planck mass and $g_*$ is
the total number of effectively relativistic degrees of freedom at the
time of freeze--out.

In order to compute the integral in eq.(\ref{eq:sigav}) reliably and
with acceptable numerical effort, we split the integral in up to three
regions. Well below and well above the resonance we use the original
form of eq.(\ref{eq:sigav}), but for $s \sim m_X^2$ we perform a
change of variable:
\begin{equation} \label{eq:trafo}
  y(s) = \arctan{\left( \frac{s-m_X^2}{m_X \Gamma_X} \right)}\,.
\end{equation}
It has been designed such that the Jacobian $ds / dy$ removes the
denominator in eq.(\ref{eq:sigav}), so that we have to compute the
integral
\begin{equation} \label{eq:poleint}
  I = \int_{y_-}^{y_+} \frac{s(s+2m^2_\psi)}{m_X  \Gamma_X} \sqrt{s-4m_\psi^2}
  K_1\left(\frac{\sqrt{s}}{T}\right)dy\,.
\end{equation}
The integration boundaries in eq.(\ref{eq:poleint}) are given by
\begin{equation} \label{eq:boundaries}
  \begin{split}
        y_-&=y(\text{max}(0.9m_X^2,4m_\psi^2))\,; \\
        y_+&=y(1.1m_X^2)\,,
  \end{split}
\end{equation}
where the argument of $y$ should be inserted in eq.(\ref{eq:trafo}).
Of course, for $4 m_\psi^2 > 1.1 m_X^2$ this is not necessary since $s$
always lies well above the resonance.

Before freeze--out dark matter is assumed to be in full kinetic and
chemical equilibrium, while for $T < T_f$, i.e. $x > x_f$, DM
annihilation dominates and reduces the dark matter abundance. This
effect is described by the annihilation integral
\begin{equation} \label{eq:J}
  J = \int_{x_f}^{\infty}\frac{\langle \sigma v \rangle}{x^2} dx \,.
\end{equation}
The present day relic density is then given by:
\begin{equation}
\Omega_\psi h^2 = \frac{1.07 \cdot 10^9 \text{GeV}^{-1}}{J g_* M_{\rm Pl}}\,.
\end{equation}
Here $\Omega_\psi$ is the mass density in DM particles (both $\psi$
and $\bar\psi$) in units of the critical density, and $h$ is the
scaled Hubble constant. Cosmological measurements imply
$\Omega_{\rm DM} h^2 \simeq 0.12$ \cite{PDG}.

\begin{figure}[hbpt]
 \centering
 \includegraphics[width=0.80\textwidth]{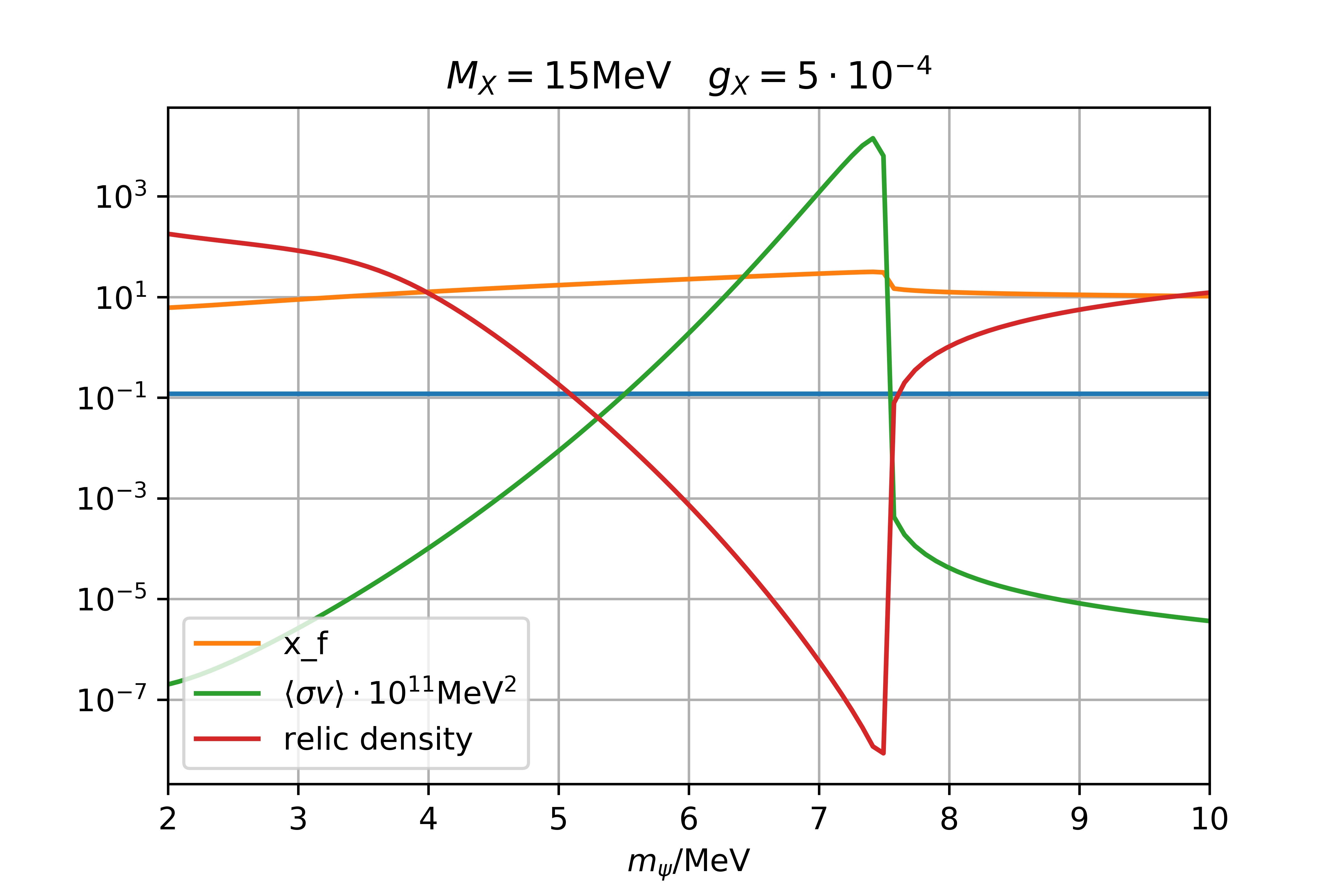}
 \caption{The predicted relic density (red), the thermal average of
   the annihilation cross section at decoupling (green) and the scaled
   inverse decoupling temperature $x_f$ (orange) as function of the DM
   mass $m_\psi$, for $m_X = 15$ MeV and gauge coupling
   $g_X=5\cdot 10^{-4}$. The blue line, representing
   $\Omega h^2 =0.12$, is the desired dark matter relic density.}
    \label{fig:relic-sample}
\end{figure}

An example of the dependence of the relic density, and related
quantities, on the DM mass $m_\psi$ is shown in
fig.~\ref{fig:relic-sample}. We see that the relic density, shown in
red, decreases quickly as $m_\psi$ approaches $m_X/2$ from below. This
is mirrored by the rise of the thermally averaged annihilation cross
section at $x = x_f$ (shown in green), and by the much slower rise of
$x_f$ (shown in orange) which depends on $\langle \sigma v \rangle$
only logarithmically. At $m_\psi = m_X/2$ the relic density shoots up
by several orders of magnitude, and then continues to rise more slowly
for yet larger DM masses. This asymmetry can be explained by the fact
that for $m_\psi < m_X/2$ the resonance $s = m_X^2$ can always be
reached for some kinetic energy of the DM particles; however, once
$m_\psi > m_X/2$ the resonance can no longer be accessed.

For the given parameters, which allow to explain the measurement of
$g_\mu - 2$, the minimum of the predicted relic density, reached for
$m_\psi$ just below $m_X/2$, is several orders of magnitude below the
desired value of $0.12$, shown in blue. Hence the correct relic
density can be obtained for two values of the DM mass, one about
$30\%$ below $m_X/2$ and the other about $1.2\%$ above $m_X/2$. It
should be noted that for the solution with $m_\psi < m_X/2$ the
thermally averaged annihilation cross section will decrease at
temperatures below $T_f$, since reduced average kinetic energy means
that $\sqrt{s}$ is usually further away from the pole; for this
solution today's annihilation cross section, which (for galaxies like
our own) should be computed at $x \sim 10^{-6}$, is considerably
smaller than $\langle \sigma v \rangle$ shown in
fig.~\ref{fig:relic-sample}. In contrast, if $m_\psi > m_X/2$ reducing
the kinetic energy moves the $\psi \bar \psi$ system closer to the
pole, so that today's value of $\langle \sigma v \rangle$
significantly exceeds that shown in fig.~\ref{fig:relic-sample}. We
will come back to this point in Sec.~4.

\subsection{Direct Dark Matter Detection}

For $m_\psi \ll 1$ GeV the most sensitive constraints come from
DM--electron scattering. In our model the relevant interaction is
described by the Lagrangian (\ref{eq:mix}), leading to a scattering
cross section of \cite{DMescattering}:
\begin{equation} \label{eq:sigma-e}
 \sigma(\psi e \rightarrow \psi e) = \frac{\mu^2_e}{\pi}
 \frac {\epsilon_A^2 e^2q_\psi^2 g_X^2} { (m_X^2 + \alpha^2 m_e^2)^2}\,,
 \quad \mu_e = \frac {m_\psi m_e} {m_\psi + m_e}\,.
\end{equation}
Several experiments have set upper bounds on this cross section:
XENON10 \cite{dmexp:XENON10}, XENON1T \cite{dmexp:XENON1T}, DarkSide50
\cite{dmexp:Darkside50} and SENSEI \cite{dmexp:SENSEI}. For a
$m_\psi \simeq 10$ MeV, the best current bound,
$\sigma(\psi e \rightarrow \psi e) \leq 0.5 \times 10^{-36}\
\text{cm}^2$, comes from the SENSEI experiment. On the other hand, for
$m_\psi \gg m_e$ (i.e. $\mu_e \simeq m_e$), $m_X^2 \gg m_e^2$ and
$\epsilon_A \simeq -g_X/70$, eq.(\ref{eq:sigma-e}) yields
\begin{equation} \label{sigma-e2}
  \sigma(\psi e \rightarrow \psi e) \simeq 6 \cdot 10^{-44} \ {\rm cm}^2
  \left( \frac {g_X} {10^{-3}} \right)^4 \left( \frac {10 \ {\rm MeV}}
    {m_X} \right)^4 q_\psi^2\,.
\end{equation}
For the parameter range of interest to us this is well below the
present and near--future sensitivity

\subsection{Big Bang Nucleosynthesis and Hubble Tension}

Both our dark matter candidate $\psi$ and the new gauge boson $X$
couple to neutrinos. If their masses are well below the muon mass they
will therefore dump their energy and their entropy into neutrinos.  If
this happens after neutrinos decouple from photons they will deliver
extra entropy and energy exclusively into neutrinos, which would
modify the history of Big Bang Nucleosynthesis (BBN) by speeding up
the expansion rate of the universe. This effect has been well studied
for the $X$ boson \cite{Neff:Bosoncase, Neff:boson2_darkmatter} and
for DM particles \cite{Neff:fermioncase} separately. Once again, we
assume that both $X$ and $\psi$ were in full thermal equilibrium with
the SM plasma at sufficiently high temperature; moreover, we assume
that the initial temperature was above $m_X$, so that the initial $X$
and $\psi/\bar\psi$ densities were comparable to those of all other
relativistic species. They then decay or annihilate into neutrinos
when they become non--relativistic.

The impact on the neutrino energy density is usually parameterized by
$N_{\rm eff}$, defined by the total relativistic energy density well
after electron--positron annihilation:
\begin{equation} \label{eq:rhor}
  \rho_{\rm rad} = \left[ 1 + N_{\rm eff} \frac{7}{8}
    \left(\frac{4}{11}\right)^{\frac{4}{3}} \right] \rho_{\gamma}\,,
\end{equation}
where $\rho_\gamma$ is the energy density in photons.  We compute
$N_{\rm eff}$ using entropy conservation, rather than solving the
Boltzmann equation. We make the following assumptions:
\begin{enumerate}
\item No chemical potential, i.e. no particle--antiparticle asymmetry
  in the neutrinos\footnote{A leptonic asymmetry of the order of the
    baryon--antibaryon asymmetry is totally negligible for our
    purposes.} or the $\psi - \bar\psi$ sector.
\item The masses of new particles are much larger than the
  recombination temperature $T_r \sim 1$ eV. For temperatures $\lsim T_r$
  the $X$ particles will then practically all have decayed away, while
  the DM particles 
  are totally non--relativistic and do not contribute to the entropy
  anymore. Of course, contributions to $N_{\rm eff}$ from $X$ decays and
  $\psi \bar \psi$ annihilation after neutrino decoupling still persist.
  
\item We assume that all neutrinos decouple at the common temperature
  $T_{\nu,D}= 2.3$ MeV. This is the decoupling temperature for $\nu_\mu$
  and $\nu_\tau$ in the SM. Since $X$ does not couple to electron
  neutrinos, and the $X$ exchange contribution to the interactions between
  neutrinos and electrons or nuclei is subdominant (see below), this should
  be a good approximation. For $T < T_{\nu, D}$ the neutrinos, $X,\, \psi$ and
  $\bar \psi$ thus form a system that is decoupled from the photons.
\end{enumerate}
  
The contributions from dark matter annihilation and $X$ boson decay are
then \cite{BBN:Neff}:
\begin{equation} \label{eq:neff}
   N_{\rm eff} = N_\nu \left[ 1 + \frac{1}{N_\nu}  \sum_{i=\psi, X} \frac{g_i}{2}
    F \left( \frac{m_i}{T_{\nu,D}} \right) \right]^{4/3}\,,
\end{equation}
where $g_\psi = 4$ (including antiparticles), $g_X = 3$ and the function
$F$ is given by
\begin{equation} \label{eq:F}
  F(x) = \frac{30}{7\pi^4} \int_x^{\infty} dy
  \frac{(4y^2-x^2)\sqrt{y^2-x^2}}{e^y \pm 1}\,;
\end{equation}
the sign $+(-)$ refers to fermion (boson) statistics. $F$ basically
describes the entropy carried by a species of massive particle at
$T = T_{\nu,D}$, which will then go into neutrinos, as explained
above. The power $4/3$ arises because we defined $N_{\rm eff}$ in
eq.(\ref{eq:rhor}) via the energy density rather than via the entropy
density; it is the energy density which determines the Hubble rate
during BBN. We checked that this simplified treatment quite accurately
reproduces numerical results in the literature \cite{Neff:Bosoncase,
  Neff:boson2_darkmatter}.

\begin{figure}[hbpt]
    \centering
\includegraphics[width=0.8\textwidth]{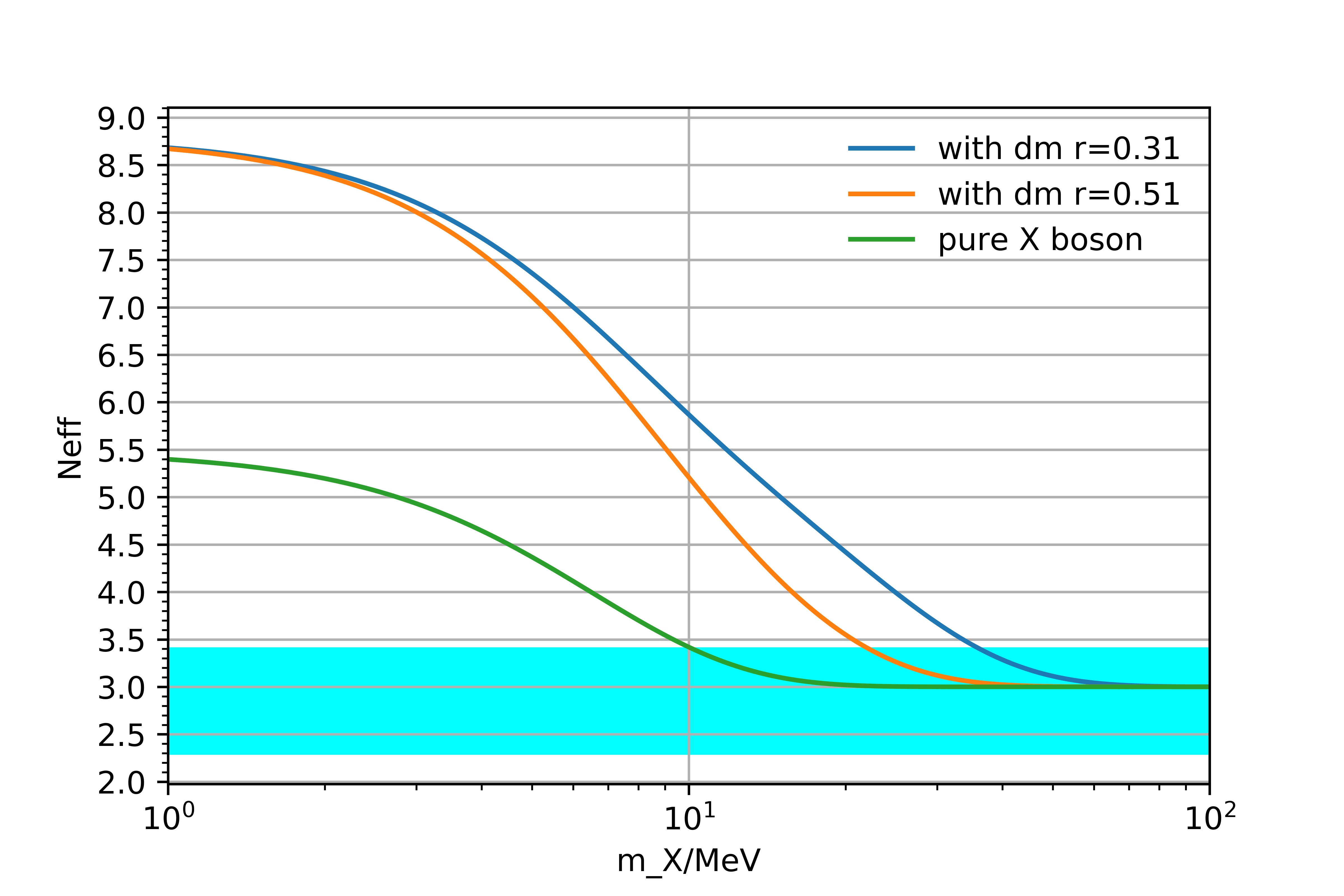}
\caption{The contribution from $X$ and $\psi$ particles to
  $N_{\rm eff}$, as function of $m_X$ with $m_\psi = r m_X$. The lower
  (green) curve shows the contribution from $X$ alone, while the
  orange and blue curves include the contribution from
  $\psi \bar \psi$ annihilation with $r = 0.51$ and $0.31$,
  respectively. The BBN allowed region \cite{PDG} is shown in cyan.}
\label{fig:Neff}
\end{figure}

Numerical results for $g_X = 5 \cdot 10^{-4}$ are shown in
fig.~\ref{fig:Neff}. For large masses the contributions to
$N_{\rm eff}$ vanish since then the decay or annihilation already
happens at $T > T_{\nu,D}$, and thus goes into the overall SM plasma;
this does not impact BBN, which happens at temperatures well below
$T_{\nu,D}$. In contrast, for masses $\lsim T_{\nu,D}$ the new
particles would clearly increase $N_{\rm eff}$ well beyond the current
bound; according to the Particle Data Group, successful BBN requires
$N_{\rm eff} \leq 3.4$ at $95\%$ confidence level.  This imposes a
lower bound on the relevant particle masses.

On the other hand, it has been shown \cite{Vagnozzi:2019ezj,
  DiValentino:2021izs} that increasing $N_{\rm eff}$ somewhat above
its SM value of $3.05$ allows to alleviate the ``Hubble tension'',
i.e. the tension between the (low) value of today's Hubble constant
derived from anisotropies of the Cosmic Microwave Background (CMB) and
other cosmological observations, and the (higher) value derived from
direct, ``local'' measurements. We consider $N_{\rm eff} \geq 3.2$ as
making a significant contribution to solving this puzzle.\footnote{The
  tension could be reduced even more if the neutrino free streaming
  length could be reduced below its SM value. However, this is only
  significant if the neutrino interaction strength at low energies,
  $g_X^2 / m_X^2$, exceeds $10^{-4} / {\rm MeV}^2$
  \cite{DiValentino:2021izs}; for $m_X \sim 20$ MeV this would give
  much too large a contribution to $g_\mu - 2$.}

\subsection{Cosmic Microwave Background}

The CMB also gives a relevant limit on dark matter annihilation into
charged particles after recombination \cite{CMB:1,CMB:2}, which would
increase the ionization fraction. This in turn would reduce the mean
free path of CMB photons, which affects the pattern of CMB
anisotropies. The resulting bound on the annihilation cross section
is given by
\begin{equation} \label{eq:CMB}
  f_{\text{eff}}(m_\psi) \frac{\langle \sigma v \rangle}{2}
  \leq 4.1 \times 10^{-28} \left( \frac{m_\psi}{\text{GeV}}\right)
  \text{cm}^3/\text{s}\,,
\end{equation}
where $f_\text{eff}$ is an ${\cal O}(1)$ parameter that describes the
energy injection efficiency into ionization. This constraint applies
to our model in two scenarios. If the dark matter mass is larger than
the muon mass, $\psi \bar{\psi} \to \mu \bar{\mu}$ becomes a main
annihilation channel for dark matter, which determines its relic
density. In this case, there is a lower bound on dark matter mass up
to several GeV, which completely excludes the sub--GeV region for
$m_\mu < m_\psi$.

If $m_\psi \lsim m _\mu$, our dark matter particles dominantly
annihilate into neutrinos, which have $f_{\rm eff} \sim 0$. The bound
(\ref{eq:CMB}) thus only applies to $\psi \bar{\psi} \to e^+ e^-$,
whose cross-section is
$\left( \frac{e\epsilon_A}{g_X} \right)^2 \approx 2\times10^{-5}$
smaller than that for $\psi \bar{\psi} \to \nu
\bar{\nu}$. Nevertheless this constraint will play a role when we will
try to explain the excess of 511 keV photons.

\subsection{White Dwarf Cooling}

White Dwarfs are created at very high temperatures. In the SM they
cool by emitting photons and neutrinos, where the latter contribution
is described by the standard weak interactions. The cooling rate can
be constrained by observing the distributions of White Dwarfs as a
function of their temperature or, equivalently, brightness. One finds
that the SM prediction reproduces the observation fairly well. This
can be used to constrain additional cooling mechanisms
\cite{Otherexp:whitedwarfscooling}.

Neutrino emission contributes to cooling mostly through the ``decay''
of plasmon quasiparticles into $\nu \bar \nu$ pairs. The rate for
this process is proportional to the square of the effective coupling
of neutrinos to the electron vector current. Since the $Z e^+e^-$
coupling in the SM is mostly axial vector, in the SM the neutrino
cooling is dominated by the emission of electron neutrinos, which also
have charged current couplings to electrons. The corresponding
effective Lagrangian can be written as
\begin{equation} \label{eq:WD-SM}
  \mathcal{L}_{\rm eff,SM} =-\frac{C_V G_F}{\sqrt{2}}(\bar{\nu} \gamma^{\mu}
  (1-\gamma_5)\nu) (\bar{e}\gamma_{\mu}e )\,,
\end{equation}
where $G_F$ is the Fermi constant and $C_V \simeq 1$
\cite{Otherexp:whitedwarfscooling}. In the \Lmt model there is an
additional contribution to this effective Lagrangian, involving $\mu$ and
$\tau$ neutrinos couplings to the electron current via eq.(\ref{eq:mix}):
\begin{equation} \label{eq:WD-us}
  \delta \mathcal{L}_{\rm eff} = -\frac{G_D}{2} q_l (\bar{\nu_l} \gamma^{\mu}
  (1-\gamma_5)  \nu_l) (\bar{e}\gamma_{\mu}e )\,,
\end{equation}
with effective coupling
\begin{equation} \label{eq:GD}
G_D =\frac{\epsilon_A g_X e}{m_X^2}\,;
\end{equation}
here $e$ is the QED coupling constant, $\epsilon_A$ has been given in
the second eq.(\ref{eq:mix}), and $q_l = 1 \ (-1)$ for
$l = \mu \ (\tau)$.

In the SM, White Dwarf cooling by neutrino emission is important only
at high temperatures. The constraint on this emission rate is therefore
not very tight; according to ref.~\cite{Otherexp:whitedwarfscooling},
the total neutrino emission rate should not exceed the SM prediction by
more than a factor of 2. Taking into account that two generations of
neutrinos contribute in eq.(\ref{eq:WD-us}), this requires $|G_D| \leq G_F$,
which in turn implies
\begin{equation} \label{eq:WDbound}
  \frac{g_X}{m_X} \leq 5.3 \cdot 10^{-4} \frac {10 \ {\rm MeV}} {m_X}\,.
\end{equation}

The same interaction can be constrained \cite{Otherexp:review} from
the analysis of $\nu e \rightarrow \nu e$ scattering, the most
sensitive data coming from the Borexino collaboration
\cite{Otherexp:Borexino}. According to \cite{Otherexp:review,
  Kaneta:2016uyt} the resulting bound is slightly weaker than that
from White Dwarf cooling, so we do not show it.

\subsection{Other Experimental Constraints}

There is already a comprehensive study on how different experiments
can constrain the parameter space of the \Lmt model
\cite{Otherexp:review}. We therefore limit ourselves to a very brief
discussion of the most relevant constraints, which are shown in
fig.~\ref{fig:g_X_m_X}.

\subsubsection{Neutrino Trident Production}

Neutrino trident production $\nu_\mu N \to \nu_\mu N \mu^+ \mu^-$ has
been shown to be a powerful tool to constrain the \Lmt model
\cite{Otherexp:neutrino_trident_production}. It gives additional
contributions via $X$ exchange between the $\nu_\mu$ and muon
lines. The cross sections reported by the CHARM-II
\cite{Otherexp:CHARMII} and CCFR \cite{Otherexp:CCFR} collaborations
are compatible with SM predictions and thus put strong limits on our
model, as shown by the red line in fig.~\ref{fig:g_X_m_X}.

\subsubsection{Coherent Elastic Neutrino Nucleus Scattering}

The coherent elastic neutrino nucleus scattering (CE$\nu$NS)
experiment constrains interactions between neutrinos and nuclei,
i.e. up and down quarks \cite{Otherexp:Coherent1}. Our \Lmt model
contributes to such process again via $X-$photon mixing, see
eq.(\ref{eq:mix}). Using data from scattering on both Cs and Ar nuclei
one can derive a strong constraint on the parameter space; see the
green curve in fig.~\ref{fig:g_X_m_X}.

\subsubsection{$X$ Production at BaBar}

The BaBar experiment searched for spin--1 bosons which couple to muons
via the process $e^+e^- \to \mu^+ \mu^- X\, ,\ \ X \to \mu^+ \mu^-$
\cite{Otherexp:Babar}. Note that they assume unit branching ratio for
$X \rightarrow \mu^+ \mu^-$ decays; in our model this branching ratio
never exceeds $50\%$. So the published bound should be interpreted as
bound on $g_X^2 \cdot {\rm Br}(X \rightarrow \mu^+ \mu^-)$. The result
is shown by the brown line in fig.~\ref{fig:g_X_m_X}.

\subsection{Summary of the Allowed Parameter Space}

\begin{figure}[ht]
\centering
\includegraphics[width=0.8\textwidth]{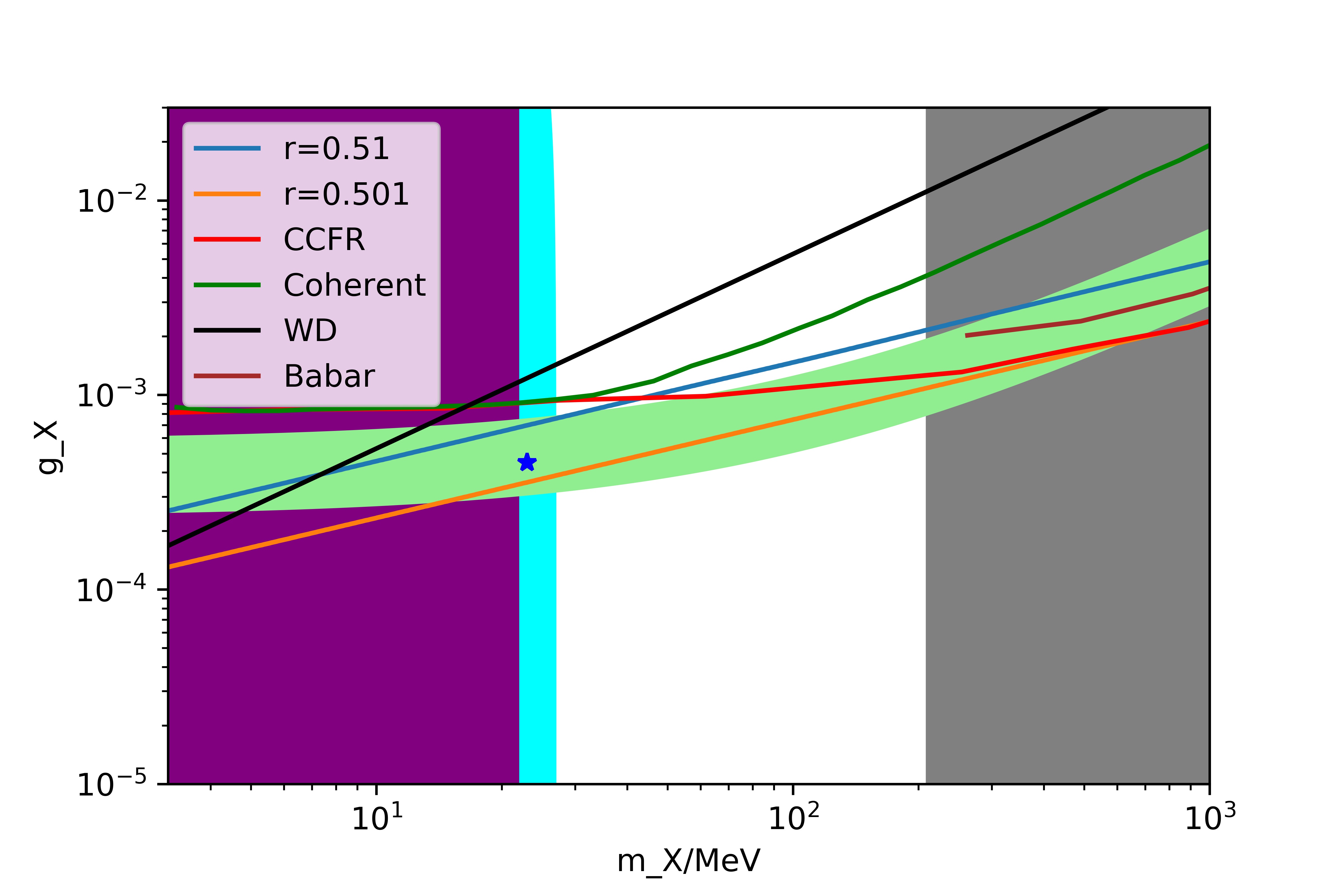}
\caption{Parameter space of the \Lmt model described by the gauge
  coupling $g_X$ and the gauge boson mass $m_X$. The purple region is
  excluded by $N_{\rm eff}$, the cyan region is favored by the Hubble
  tension, the gray region is excluded by the CMB, and the light green
  region allows to explain the measurement of $g_\mu - 2$ within one
  standard deviation. The black, dark green, red, and brown lines are
  upper bounds on $g_X$ from White Dwarf (WD) cooling, neutrino trident
  production (CCFR), neutrino nucleus coherent scattering, and the
  BaBar experiment. Parameters on the blue (orange) line reproduce the
  correct thermal relic dark matter density for
  $m_\psi/m_X =0.51 \ (0.501)$ and charge $q_\psi=1$. The blue star is
  the benchmark point used to study the 511 keV line.}
\label{fig:g_X_m_X}
\end{figure}

The constraints discussed in this Section are summarized by
fig.~\ref{fig:g_X_m_X}. The BBN constraint $N_{\rm eff} < 3.4$
requires $m_X \gsim 20$ MeV, the precise value depending on the ratio
$m_\psi/m_X$, as shown in fig.~\ref{fig:Neff}. On the other hand, the
Hubble tension can be alleviated only for $N_{\rm eff} \geq 3.2$,
which requires $m_X \lsim 27$ MeV. These two constraints thus
essentially fix the mass of the new gauge boson. The measurement of
$g_\mu -2$ can then be used to determine the new gauge coupling $g_X$,
as indicated by the light green region. Finally, the requirement that
$\psi$ has the correct thermal relic density in minimal cosmology
determines $m_\psi$, or $m_\psi / m_X$, as shown in
fig.~\ref{fig:relic-sample}.

So for a very restricted region of parameter space our model can
already (help to) solve three puzzles: the existence of cosmological
dark matter, the discrepancy between the SM prediction for and the
measurement of $g_\mu - 2$, and the discrepancy between
``cosmological'' and ``local'' measurements of the Hubble constant. We
will now show that this model might explain yet another long--standing
puzzle.

\section{511keV Line}

Starting in the 1970's a large flux of photons with energy of $511$
keV from the center of our galaxy has been observed \cite{511:1970_1,
  511:1970_2, 511:1970_3, 511:INTEGRAL, 511:COSI}. This sharp line can
obviously be interpreted as originating from the annihilation of
non--relativistic $e^+e^-$ pairs. The challenge is to explain the
rather high density of positrons required to reproduce the
measurements.

It has been pointed out more than 15 years ago that dark matter
annihilation into $e^+ e^-$ pairs could be responsible for this
signal \cite{511:DM_1, 511:DM_2}. This requires an annihilation cross
section of
\begin{equation} \label{eq:sig-ee}
  10^{-3} \text{fb} \leq \langle \sigma(\psi \bar \psi \rightarrow
  e^+ e^-)v\rangle \cdot \left( \frac{m_\psi}{1\ \text{MeV}}\right)^{-2}/2
    \leq 1\text{fb}
\end{equation}
for Dirac DM with $n_\psi = n_{\bar\psi}$; the averaging is now over
the DM velocity distribution in the region of the galactic center,
where typically $v \sim 10^{-3}$ (in natural units). Here we use the
same conservative range as in \cite{511:DM_Manuel}, which is expanded
by one order of magnitude compared with \cite{511:DM_2}. Note that the
upper end of the range (\ref{eq:sig-ee}) is excluded by CMB
constraints: taking $f_{\text{eff}}(10\ \text{MeV})= 0.9$
\cite{CMB:1}, eq.(\ref{eq:CMB}) gives
\begin{equation} \label{eq:CMBbound}
  \frac{\langle \sigma v \rangle}{2} \leq 0.015 \left(
    \frac{m_\psi}{1 \ \text{MeV}}\right) \text{fb}\,.
\end{equation}
Since this bound depends linearly on the DM mass while the required
cross-section grows quadratically with $m_\psi$,\footnote{The
  difference comes about because the CMB restricts the {\em energy}
  injected in form of $e^+e^-$ pairs, whereas the flux of $511$ keV
  photons is proportional to the {\em number} of such pairs.} it
prevents DM particles with mass above $15$ MeV from explaining the
$511$ keV line. However, upper bounds on the flux of MeV photons from
the galactic center in any case imply \cite{Beacom:2004pe,
  Beacom:2005qv} $m_\psi < 3$ MeV if DM annihilation alone explains
the positron flux. This stringent constraint can be relaxed by
reducing the dark matter contribution. We estimate that the
annihilation of $10$ MeV DM particles can still contribute $30$ to
$40\%$ of the 511 keV flux. The remainder would then have to originate
from astrophysical processes producing at most mildly relativistic
positrons; note that the upper bound of $3$ MeV, which originates from
positron annihilation in flight \cite{Beacom:2005qv}, really refers to
the positron injection energy, which equals $m_\psi$ for positrons
from $\psi \bar \psi \rightarrow e^+e^-$ annihilation.

Recall that the DM coupling to $e^+e^-$ is loop suppressed in our model;
the cross section for annihilation into $e^+e^-$ pairs can thus be
estimated as
\begin{equation} \label{eq:sig-ee-us}
  \langle \sigma(\psi \bar \psi \rightarrow e^+e^-) v \rangle \approx
  \left( \frac{e}{70} \right)^2 \langle
  \sigma(\psi \bar \psi \rightarrow \nu \bar{\nu}) v \rangle
  \approx 2\times 10^{-5} \langle
  \sigma(\psi \bar \psi \rightarrow \nu \bar{\nu}) v \rangle\,.
\end{equation}
In the non--relativistic limit appropriate for DM annihilation in
today's universe the cross section of eq.(\ref{eq:sigma}) can be
simplified. Using
\begin{equation} \label{eq:s}
  s \approx 4m^2_\psi + m^2_\psi v^2_{\rm rel}\,,
\end{equation}
where $v_{\rm rel}$ denotes the relative velocity between annihilating
DM particles in the center--of--mass frame, and defining
\begin{equation} \label{eq:defs}
 r = m_\psi/m_X\,;  \quad \delta=4r^2-1\,; \quad \gamma_X = \Gamma_X / m_X\,,
\end{equation}
the annihilation cross section can be written as
\begin{equation}
  \sigma (\psi \bar \psi \rightarrow \nu_l \bar{\nu}_l) v
  \approx  \frac{q^2_\psi g^4_X}{8\pi m_X^2}
    \frac{1}{(\delta+v_{\rm rel}^2/4)^2+\gamma^2_X} \,.
\end{equation}
Due to the loop suppression in eq.(\ref{eq:sig-ee-us}) we need today's
$\psi \bar\psi$ annihilation cross section to be significantly larger
than that at decoupling if DM annihilation is to contribute
significantly to the positron flux. As explained at the end of
Sec.~3.1 we thus need $r$ to be slightly larger than $0.5$, so that
the cross section is Breit--Wigner enhanced with an unphysical pole
\cite{Breit-WignerEnhancement, Breit-WignerEnhancement2}. The
enhancement of today's averaged annihilation cross section over that
at decoupling is called the ``boost factor'' (BF); it is determined by
$\delta$ and $\gamma_X$:
\begin{equation} \label{eq:BF}
\text{BF} = \frac{\max[\delta,\gamma_X]^{-1}}{10}\,.
\end{equation}
In the interesting region of parameter space $\gamma_X \sim 10^{-8}$
is always much smaller than $\delta$. For $r > 0.5$ we need an
averaged DM annihilation cross at decoupling of about $1$
pb\footnote{For $r < 0.5$, $\langle \sigma v \rangle$ at decoupling
  actually needs to be closer to $50$ pb, because in that case
  $\langle \sigma v \rangle$ drops very quickly with increasing $x$,
  reducing the annihilation integral $J$; this effect would also make
  DM annihilation in today's universe unobservable for $r < 0.5$.};
this corresponds to $v^2_{\rm rel}/4 \simeq 0.1$. Today
$v_{\rm rel}^2/4 \lsim 10^{-6}$ in our galaxy; for $\delta > 10^{-6}$
today's averaged cross section is then approximately given by
\begin{equation} \label{eq:sig-now}
  \langle \sigma v \rangle_{\rm now} \approx \frac{\delta^{-1}}{10}\
  \text{pb}\,.
\end{equation}

Of course, $\delta$ cannot be chosen arbitrarily; as we saw in
Sec.~3.1, it is determined by the requirement that the DM relic
density comes out right. In our previous analysis we had set the DM
charge $q_\psi = 1$. A smaller DM charge will force the DM mass to be
closer to the pole, giving a smaller value of $\delta$. We can
therefore increase today's annihilation cross section into $e^+e^-$
pairs by reducing $q_\psi$; of course, we still need to take care not
to violate the CMB constraint.

\begin{figure}[hbpt]
   \centering
\includegraphics[width=0.8\textwidth]{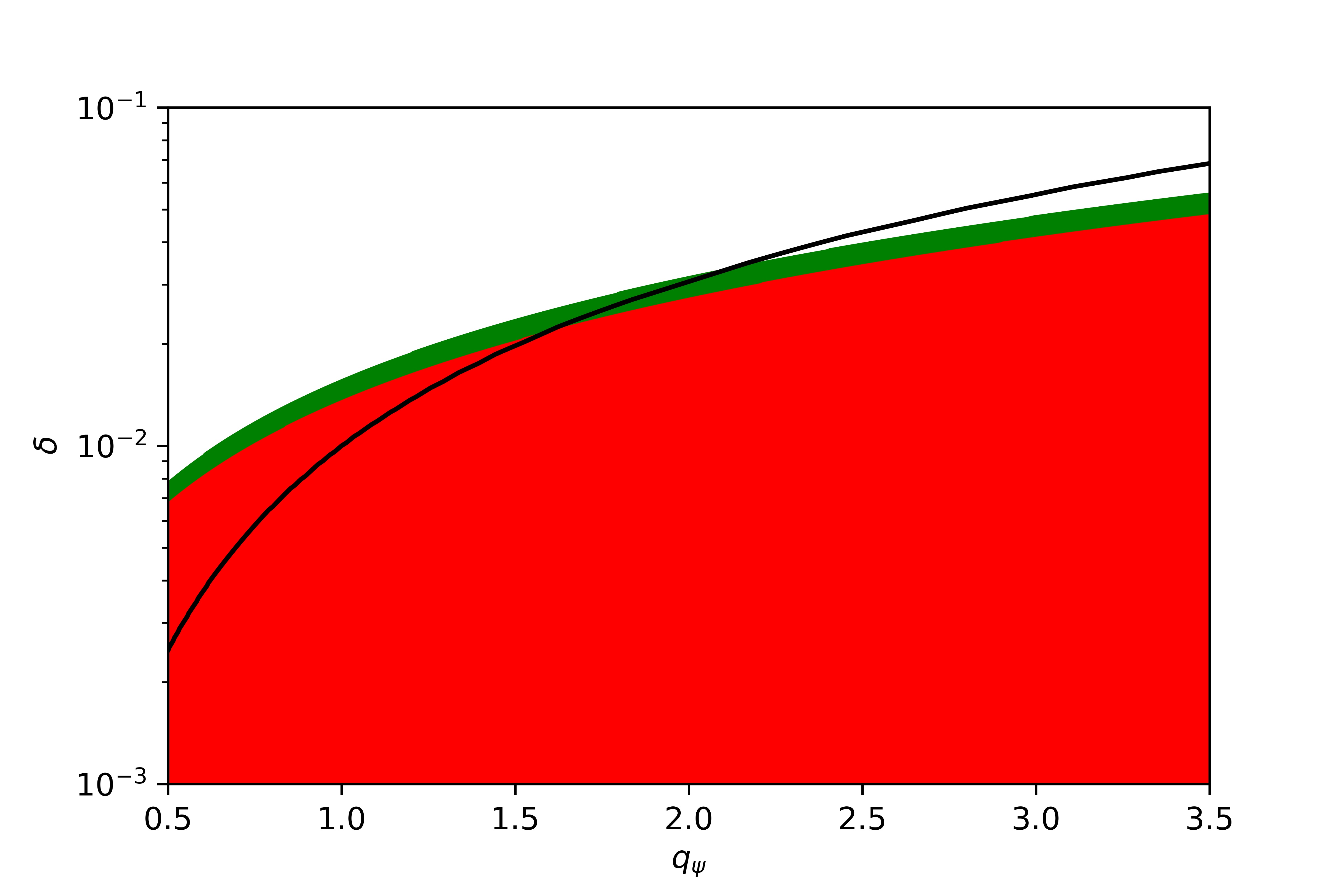}

\caption{The parameter space spanned by DM charge $q_\psi$ and
  $\delta = 4 m_\psi^2/m_X^2 - 1$, for gauge coupling
  $g_X=4.5 \times 10^{-4}$ and $m_X=23$ MeV. On the black line the
  thermal relic density in minimal cosmology has the desired value. In
  the green region DM annihilation might explain the $511$ keV line,
  while the red region is excluded by the CMB constraint
  (\ref{eq:CMBbound}).}
    \label{fig:qpsi}
\end{figure}

Fig.~\ref{fig:qpsi} shows numerical results in the $q_\psi - \delta$
plane.  We see that DM annihilation in our model might explain the
excess of $511$ keV photons if
$1.5 \lsim q_\psi \lsim 2.2$\footnote{This range is quite sensitive to
  gauge coupling $g_X$: for $g_X = 5.4\cdot 10^{-4}$,
  $1.1 \lesssim q_\psi \lesssim 1.5$ is preferred while for
  $g_X = 3.2\cdot 10^{-4}$, $ 3.1 \lesssim q_\psi \lesssim 4.1$ is
  needed. These values bracket the $2\sigma$ range of $g_\mu - 2$.};
for even smaller values of $q_\psi$ the cross section for annihilation
into $e^+e^-$ pairs violates the CMB constraint
(\ref{eq:CMBbound}). Due this constraint DM annihilation can probably
produce a sizable contribution to the required positron flux only if
the DM density peaks rather strongly close to the galactic
center. Note also that the constraints shown in Fig.~\ref{fig:g_X_m_X}
do not depend on $q_\psi$. The DM charge thus offers a handle to tune
the contribution to the flux of $511$ keV photons from DM annihilation.

\section{Future Tests}

Future tests of this model have also been discussed in
ref.~\cite{Otherexp:review}. We therefore limit ourselves to two
remarks.

First, for parameters that allow to explain the $511$ keV line, the
discussion of the previous Section shows that our model predicts a
large annihilation cross section into neutrinos in today's universe,
$\langle \sigma(\psi \bar\psi \rightarrow \nu_l \bar{\nu}_l)v
\rangle_{\rm now} \approx 10^{-24}-10^{-25} \
\text{cm}^3/\text{s}$. According to estimates in
\cite{Neutrinoexp:Sensitivity} the corresponding flux of neutrinos
with $E_\nu = m_\psi \simeq 10$ MeV should be easily detectable at
next generation neutrino experiments like JUNO \cite{neutrinoexp:JUNO}
through neutrino--electron scattering.

Secondly, in the entire range $m_X \lsim 200$ MeV where the model can
simultaneously produce a good thermal DM candidate and explain the
measurement of $g_\mu - 2$ without violating existing constraints, it
might also be testable through the process $e^+ e^- \to X \gamma$
followed by invisible decays of the $X$ boson
\cite{511:DM_Manuel}. The differential cross section reads:
\begin{equation} \label{eq:belle1}
 \frac{d\sigma}{d\cos{\theta}} = \frac{\alpha (e \epsilon_A)^2}{s (s-m_X^2)}
 \left[ 
    \frac{s^2+m_X^4}{\sin{\theta}^2} - \frac{(s-m_X^2)^2}{2}
    \right]\,,
\end{equation}
where $\alpha$ is the fine structure constant and $\theta$ is the
angle between the photon and the beam direction in the
center--of--mass frame. Requiring $\abs{\cos{\theta}} < 0.985$ in
order to make sure that the photon can be detected and taking
$m_X^2 \ll s$ leads to a cross section of about
\begin{equation} \label{eq:belle2}
  \sigma \simeq 20 \text{ab} \left(\frac{g_X}{10^{-3}}\right)^2
  \frac{(10\ \text{GeV})^2}{s}\,.
\end{equation}
The signature is a photon with center of mass energy
$E_\gamma = (s- m_X^2)/(2 \sqrt{s}) \simeq \sqrt{s}/2$, without any
other detectable particles. This has very little physics background
from $e^+e^- \rightarrow \nu \bar \nu \gamma$, the cross section for
which scales like $\alpha G_F^2 s$. Currently the $B-$factory Belle-II
operates at $\sqrt{s} \simeq 10$ GeV; its goal is to achieve an
integrated luminosity of $50$ ab$^{-1}$ \cite{Belle-II:2010dht},
leading to hundreds of signal events for $g_X = 5 \cdot
10^{-4}$. Similarly, the proposed super tau--charm factory in China
would operate at $\sqrt{s} \simeq 4$ GeV with integrated luminosity
around $2$ ab$^{-1}$ per year; this would yield a similar number of signal
events with even less physics background.

\section{Summary and Conclusions}

In this paper we showed that the \Lmt model with a Dirac dark matter
particle $\psi$ can be used to explain four different
observations. The measurement of $g_\mu-2$, which exceeds the SM
prediction by $4.2$ standard deviations, can be explained through the
exchange of the new light gauge boson $X$ for \Lmt coupling
$g_X \simeq 5 \cdot 10^{-4}$. For this value of $g_X$ the thermal dark
matter relic density comes out correct for DM charge $q_\psi \simeq 1$
if $m_\psi \simeq m_X/3$ or if $m_\psi$ is just above
$m_X/2$. Moreover, if $m_X \simeq 20$ MeV the decoupling of $X$ and
$\psi$ increases the radiation content of the universe by
$\delta N_{\rm eff} \lsim 0.4$, which respects BBN constraints but
allows to alleviate the tension between cosmological and ``local''
measurements of the Hubble constant. Finally, for the same range of
$m_X$ and $m_\psi$ just above $m_X/2$,
$\psi \bar \psi \rightarrow e^+ e^-$ annihilation might help to
explain the excess of $511$ keV photons from the center of our
galaxy. This part of parameter space is compatible with all other
experimental and astrophysical constraints. Moreover, there are two
other, less well established, observations which might prefer this
scenario over the SM: our $X$ boson will slightly speed up the cooling
of hot White Dwarfs via the emission of $\nu_\mu$ and $\nu_\tau$
pairs; according to ref.~\cite{Isern:2018uce} there is indeed some
evidence for additional White Dwarf cooling (which is interpreted in
terms of axion emission in that work). Secondly, this range of $m_X$
also appears to be slightly favored by IceCube 6 years shower data
relative to the SM--only hypothesis \cite{Intro:ICEcube}.

As discussed in ref.~\cite{Otherexp:review} this model can be tested in
future beam dump experiments. We point out that the entire parameter range
that can explain both $g_\mu - 2$ and the DM relic density should be
testable through single photon searches at super flavor factories. This
requires the implementation of a single photon trigger; we urge our
experimental colleagues in the BELLE--II collaboration, as well as
those working on the design of detectors for future tau--charm factories,
to consider this. Moreover, the choice of parameters that allows to explain
the excess of $511$ keV photons leads to a large flux of $\nu_\mu$ and
$\nu_\tau$ with energy $E_\nu = m_\psi \simeq 10$ MeV, which should be
detectable by future neutrino experiments like DUNE.

\subsection*{Acknowledgments:} We thank John Beacom for bringing
ref.\cite{Beacom:2005qv} to our attention.

\vspace*{5mm} \noindent
{\bf Note added:} In the final stages of our work
ref.~\cite{L23newpaper} appeared, which has considerable overlap with
this work. Our results in general agree with their's. However, they do not
discuss white dwarf constraints, the $511$ keV excess, and possible signals
at low--energy $e^+e^-$ colliders.


\bibliography{mybibfile}

\end{document}